\documentclass[useAMS,usenatbib]{mn2e}

\usepackage{graphicx}
\usepackage{amssymb}
\usepackage{amstext}
\usepackage{url}
\usepackage{amsmath}
\usepackage{setspace}
\usepackage{xtab}
\usepackage{textcomp}
\usepackage{booktabs,array}
\usepackage{epstopdf}
\usepackage{color}
\usepackage{colortbl}
\usepackage{tabularx}
\usepackage{gensymb}
\usepackage{subfigure}
\usepackage{placeins}
\usepackage{float}
\usepackage[normalem]{ulem}
\usepackage{rotating}
\usepackage{longtable}
\usepackage{capt-of}
\usepackage{scalefnt}

\newcommand{\Teff}{\ensuremath{T_{\mathrm{eff}}}}
\newcommand{\logg}{\ensuremath{\log g}}

\newcommand{\cd}{d\ensuremath{^{-1}}}
\newcommand{\kms}{\ensuremath{\textrm{km\,s}^{-1}}}
\newcommand{\vsini}{\ensuremath{v\sin i}}

\newcommand{\msun}{\ensuremath{\textrm{M}_{\odot}}}
\newcommand{\rsun}{\ensuremath{\textrm{R}_{\odot}}}

\title[Mode Identification of Gamma Doradus]{Frequency and mode identification of $\gamma$ Doradus from photometric and  spectroscopic observations\thanks{This paper  includes data taken at the University of Canterbury Mount John Observatory, New Zealand, the Sutherland South African Astronomical Observatory (SAAO), Cape Town Observatory operated by SAAO and Cerro Tololo Inter-American Observatory, Chile.}} 

\author[E. Brunsden, K. R. Pollard, D. J. Wright, P. De Cat, P. L. Cottrell]{E. Brunsden$^{1,2}$\thanks{E-mail: emily.brunsden@gmail.com}, K. R. Pollard$^{2}$, D. J. Wright$^{3}$, P. De Cat$^{4}$, P. L. Cottrell$^{5}$\\
$^{1}$Department of Physics, University of York, Heslington, York, YO10 5DD, UK\\ 
$^{2}$Department of Physics and Astronomy, University of Canterbury, Private Bag 4800, Christchurch, New Zealand\\
$^{3}$Department of Astrophysics, University of New South Wales, Sydney, NSW 2052 Australia\\
$^{4}$Royal Observatory of Belgium, Ringlaan 3, 1180 Brussel, Belgium\\
$^{5}$Monash Centre for Astrophysics, School of Physics and Astronomy, Monash University, Victoria 3800, Australia}

\begin{document}
\date{ }

\pagerange{\pageref{firstpage}--\pageref{lastpage}} \pubyear{2017}

\maketitle

\label{firstpage}

\begin{abstract} The prototype star for the $\gamma$ Doradus class of pulsating variables was studied employing photometric and spectroscopic observations to determine the frequencies and modes of pulsation. The four frequencies found are self-consistent between the observation types and almost identical to those found in previous studies ($1.3641$ d$^{-1}$,$1.8783$ d$^{-1}$, $1.4742$ d$^{-1}$ and $1.3209$ d$^{-1}$). Three of the frequencies are classified as $l,m=(1,1)$ pulsations and the other is ambiguous between $l,m=(2,0)$ and $(2,-2)$ modes. Two frequencies are shown to be stable over twenty years since their first identification. The agreement in ground-based work makes this star an excellent calibrator between high-precision photometry and spectroscopy with the upcoming TESS observations and a potential standard for continued asteroseismic modelling.
\end{abstract}


\section{Introduction}

Stars showing g-mode pulsations, such as $\gamma$ Doradus ($\gamma$ Dor) stars are important targets for asteroseismic investigation. Revealing the interior of stars through the determination of properties such as convection, opacity, rotation, and ionization has profound impacts on our understanding of stellar evolution. The $\gamma$ Dor stars are a key class where the high-order non-radial g-mode pulsations propagate deep into the star, allowing us to probe large regions of their interior.

The excitation mechanism for $\gamma$ Dor stars is explained by convective flux blocking \citep{2000ApJ...542L..57G,2005AandA...435..927D}, although turbulent convection has also been considered as an excitation of oscillations \citep{2015AandA...584L...2G, 2016MNRAS.457.3163X}.

The class has a localised position on the Hertzsprung-Russell diagram, on or near the main sequence at the red edge of the classical instability strip, which gives each $\gamma$ Dor star a similar overall structure. $\gamma$ Dor stars are found in the transition region from convective cores and radiative envelopes to radiative cores and convective envelopes, making them useful tools for probing this boundary to improve theoretical modelling. Intensive efforts are being made to understand $\gamma$ Dor asteroseismology and to determine their internal conditions, such as the sizes of their convection zones and their internal rotation rates (some recent examples include \citeauthor{2015EPJWC.10101005B}, (\citeyear{2015EPJWC.10101005B}), \citeauthor{2016AandA...593A.120V}, (\citeyear{2016AandA...593A.120V}), \citeauthor{2016AandA...592A.116S}, (\citeyear{2016AandA...592A.116S})). 

Mode identifications of $\gamma$ Dor stars are required to provide inputs and to test theoretical stellar models. The list of $\gamma$ Dor stars with partial or complete mode identifications for even one frequency is short (only around eight stars with full geometric mode identifications having been published to date). Precision photometric methods still generally lack spectroscopic verification and in some cases show stark disagreement (e.g. \citeauthor{2008AandA...489.1213U}, \citeyear{2008AandA...489.1213U}, \citeauthor{2015MNRAS.447.2970B}, \citeyear{2015MNRAS.447.2970B}). Using both photometric and spectroscopic data in a complementary way allows us to test our frequency and mode identification methods.

The prototype star of the class, $\gamma$ Doradus itself (HD\,27290, HR\,1338, HIP\,19893) is a bright ($V = 4.20$), F1 V \citep{2006AJ....132..161G} star which has been observed for many decades as a known variable. It was defined as the class prototype in 1999 \citep{1999PASP..111..840K}. This paper details the analysis of extensive sets of photometric and spectroscopic data obtained over a 20 year period. Section \ref{obsgam} outlines these data and section \ref{amet} their treatment. Section \ref{full4} and Section \ref{gam_modeid} describe the frequency and mode identification. A  discussion of all the results and a concluding statement is found in Section \ref{disc4} and Section \ref{10disc}, respectively.


\section{Observations}\label{obsgam}

One of the most studied $\gamma$ Dor class members, new observations are presented here of the prototype star. Stellar parameters have been previously measured for $\gamma$ Doradus. Restricting reports to results no older
than 1995, Table~\ref{10param} lists recent values found for \Teff, \logg\ and \vsini. The results show consistent findings for \Teff\ of around $7150$~K and
\logg\ near $4.1$. These values are typical for $\gamma$ Dor group members (see, for example, \citeauthor{2017MNRAS.470.4408K}, \citeyear{2017MNRAS.470.4408K}).

\begin{table}\caption[Stellar parameters found for $\gamma$ Doradus in previous studies.]{Stellar parameters found for $\gamma$ Doradus in previous
studies. Entries with $\star$ are values from spectroscopy and entries with $\dagger$ are from Str\"omgren photometry.}\label{10param}
\begin{center}
\begin{tabular}{ccccc}
\toprule
\Teff\ & \logg\ & \vsini\ & Ref \\
 &  & (\kms)& \\
\midrule
7164& &59.5$\pm$3& \citet{2012AandA...542A.116A}$\star$\\
7060& 3.97&  & \citet{2006AJ....132..161G}$\star$ \\
7120& 4.25 &  & \citet{Dupret2005}$\dagger$\\
7202& 4.23 &  & \citet{Dupret2005}$\dagger$\\
&  & 62 & \citet{1999PASP..111..840K}$\star$\\
7180&  &  & \citet{1995AandAS..110..553S}$\star$\\
\bottomrule
\end{tabular}
\end{center}  
\end{table}

$\gamma$ Doradus has been a known variable star since observations by \citet{1960MNSSA..19...56C} and \citet{1963MNSSA..22...65C}.A summary of the frequencies previously found in
$\gamma$ Doradus is given in Table~\ref{10fcomplit1}. The
first
periods were found in Str\"omgren photometry and published in \citet{1992Obs...112...53C} and confirmed in \citet{1994MNRAS.267..103B} also using
Str\"omgren photometry.
Further observations using multi-site Str\"omgren and Johnson filter photometry again supported these periods and found evidence for a third
\citep{1994MNRAS.270..905B}. In multi-site spectroscopy
analysed in the same study, $f_{l1}$ and $f_{l2}$ were detected in the radial velocity
measurements. Mode identification from the spectroscopic line profile variations \citep{1986MNRAS.219..111B,1986MNRAS.220..647B,1987MNRAS.224...41B} classified
$f_{l1}$, $f_{l2}$ and $f_{l3}$ as $(l,m) = (3,3), (1,1)$ and $(1,1)$ respectively \citep{1996MNRAS.281.1315B}. These modes were found with an
inclination of $70\degree$. Modelling by \citet{Dupret2005} identified $f_{l1}$ and $f_{l2}$ to be both well-fitted by $l=1$ modes using
time-dependent convection models on the photometric data. Observations with the Solar Mass Ejection Imager (\textsc{smei}) aboard the Coriolis satellite found
the
previously known $f_{l1}$, $f_{l2}$ and $f_{l3}$, then further identified $f_{l4}$ and $f_{l5}$ \citep{2008AandA...492..167T}.
The
authors also note the occurrence of a frequency at $2.743$~\cd, identifying it as a combination frequency of $f_{l2}$ and $f_{l5}$.  

\begin{table}\caption[Previously identified frequencies in $\gamma$
Doradus.]{Previously identified frequencies in $\gamma$
Doradus. The formal uncertainty in the last digit is displayed in
parentheses.}\label{10fcomplit1}
\begin{center}
\begin{tabular}{cccccc}
\toprule
\textsc{id} & Phot.\footnotemark[1]& Phot.\footnotemark[2] & Phot.\footnotemark[3] & Spec.\footnotemark[4] & Phot.\footnotemark[5]\\
& (\cd)& (\cd)& (\cd)& (\cd)& (\cd)\\
\midrule
$f_{l1}$& 1.321 &1.32099 & 1.32098 (2)	& 1.365 (2) 	& 1.32093 (2) \\
$f_{l2}$& 1.364 &1.36353 & 1.36354 (2)	&  		& 1.36351 (2) \\
$f_{l3}$&  	&	  &  1.475 (1) 	&  		& 1.471 (1) \\
$f_{l4}$&  	&	  & 		&  		&  1.39 (5)\\
$f_{l5}$&	&	  &		&		&1.87 (3)		\\
\bottomrule
\end{tabular}
\end{center}
\begin{center}
\footnotemark[1]\footnotesize{\citet{1992Obs...112...53C}},\footnotemark[2]\footnotesize{\citet{1994MNRAS.267..103B}},\footnotemark[3]\footnotesize{
\citet{1994MNRAS.270..905B}}, \footnotemark[4]\footnotesize{\citet{1996MNRAS.281.1315B}},
\footnotemark[5]\footnotesize{\citet{2008AandA...492..167T}}
\end{center}
\end{table}
Photometric data were supplied by L. Balona and have been previously analysed for frequencies
\citep{1992Obs...112...53C,1994MNRAS.267..103B,1994MNRAS.270..905B,1996MNRAS.281.1315B}. These data were taken at a
variety of sites and in several photometric filters, the details of which are described in Table \ref{2allphot}.

\begin{figure}
\centering
   \includegraphics[width=0.5\textwidth]{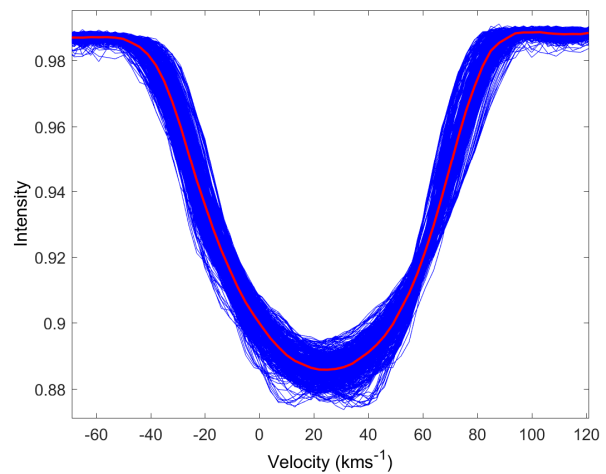}\caption[Line profiles of $613$ observations of $\gamma$ Doradus.]{Line profiles of $613$
observations of $\gamma$ Doradus. The mean profile is over-plotted in red. This star shows large pulsational variation from observation to
observation, particularly near the centre of the profile.}\label{gam_lpv}
 \end{figure}

Spectra from the University of Canterbury Mt John Observatory (\textsc{ucmjo}) in New Zealand were collected using the $1$m McLellan telescope with the fibre-fed High Efficiency and Resolution Canterbury University Large \'Echelle Spectrograph (\textsc{hercules}) with a resolving power of R $= 50000$ operating over a range of $3800~$\AA\ to $8000~$\AA\ \citep{2002ExA....13...59H}. We observed $\gamma$ Doradus over $15$ months from June 2011 to August 2012 at \textsc{ucmjo}, during which time we obtained $625$ observations. Typically we took $10-20$ observations in a single night. The sampling of the data can be seen in the window function in Figure~\ref{gam_pbpfou}. We produced a total of $613$ line profiles for analysis which show clear variation (Figure~\ref{gam_lpv}). 

\begin{table}\caption[Summary of photometric observations.]{Summary of photometric observations of $\gamma$ Doradus with number of observations, dates, total time span ($\Delta$ T), filter sets used (with number of filters)
and
references to the original publications. }\label{2allphot}
\begin{center}
\begin{tabular}{l c c c c c}
\toprule
 N of &  Dates & $\Delta$ T & Filters &  Ref. \\
 Obs. & Observed  & (days) & (number) & \\
\midrule
\\
\multicolumn{5}{c}{\textbf{\textsc{suth}} photometry}\\
447 & Jan 1993-Nov 1994 & 693 & Str\"om. (4)& \footnotemark[1],\footnotemark[2],\footnotemark[3] \\
 175 & Nov 1989-Mar 1990 & 132 & John. \textit{V} (1)& \footnotemark[4] \\
\\
\multicolumn{5}{c}{\textbf{\textsc{cape}} photometry}\\
 167 & Oct 1981-Mar 1991 & 3427 & Str\"om. (4)& \footnotemark[4]\\
 151 & Dec 1993-Jan 1994 & 80 & John. \textit{B} \textit{V}& \footnotemark[2] \\
  &  &  & \textsc{ddo} 45,48 (4)&  \\
\\
\multicolumn{5}{c}{\textbf{\textsc{ctio}} photometry}\\
 400 & Jan 1994-Nov 1994 & 325 & John. \textit{V} (1)& \footnotemark[2],\footnotemark[3]\\
\bottomrule
\end{tabular}\end{center}

\footnotemark[1]\footnotesize{\citet{1994MNRAS.267..103B}}, \footnotemark[2]\footnotesize{\citet{1994MNRAS.270..905B}},\\
\footnotemark[3]\footnotesize{\citet{1996MNRAS.281.1315B}}, \footnotemark[4]\footnotesize{\citet{1992Obs...112...53C}.}
\\

\end{table}

\section{Analysis Methods}\label{amet}

We performed time-series analysis using \textsc{famias} \citep{2008CoAst.157..387Z}, applied as in \citet{2006AandA...455..235Z} and in {\sevensize SIGSPEC} \citep{2007AandA...467.1353R}. For spectroscopic data this entailed an analysis of
the variations in the representative line profiles which identified the frequencies present. For photometric data, we analysed white light or multi-colour photometry for periodic signals.

The amplitude ratio method \citep{1979MNRAS.189..649B,1988ApandSS.140..255W,1994AandA...291..143C,2002AandA...392..151D,2003A&A...398..677D} uses calculations of
the ratios between the amplitudes and differences in phases of different filters in a multi-coloured photometric system to determine $l$ values of the modes. Typically this method constrains the $l$ values rather than giving unique solutions. We used the photometric analysis toolbox in {\sevensize FAMIAS} \citep{2008CoAst.157..387Z}.

\subsection{Delta Function Cross Correlation}
To maximise the signal obtained from the line profile variations, we used a $\delta$-function cross-correlation technique. To prepare the normalised spectra, telluric regions and the broad
hydrogen lines, H$\alpha$, H$\beta$ and H$\gamma$, are removed.  

A line list is generated by Synspec \citep{2011ascl.soft09022H}, and from this list, very weak lines (equivalent widths less than 5.0 m$\dot{\mathrm{A}}$) are removed. Cross-
correlation of the remaining lines (typically 2000 to 5000) is done using standard cross-correlation techniques and then using the $\delta$-function template. The scaled
$\delta$-function cross-correlation technique (\citeauthor{NewEntry3}, \citeyear{NewEntry3}, building on \citeauthor{2007CoAst.150..135W}, \citeyear{2007CoAst.150..135W}) cross-correlates the spectra with a template of $\delta$ functions at the correct wavelengths and line depths. Cross-correlation produces a representative spectral line, or zero-point line profile, for each of the observations. The cross-correlation process gives
high signal-to-noise representative line profiles for the star. 

The pixel-by-pixel method \citep{2000ASPC..210..138M} is a line-profile variation method that
looks at the movement of each individual pixel in a spectral absorption line and analyses
them as a time series with the calculation of a Fourier spectrum. Each pixel is then used
to create an average Fourier spectrum of the frequencies present in the line profile.

We used the Fourier spectra for the Pixel-by-Pixel (\textsc{pbp}) technique applied to the representative line profiles to identify frequencies. We then analysed these frequencies to determine mode identifications using the Fourier Parameter Fit method \citep{2009AandA...497..827Z}. 

\section{Frequency Analysis}\label{full4}

We performed frequency analysis on both spectroscopic and photometric data in \textsc{famias} and SigSpec. We restricted the analysis to frequencies within an extended $\gamma$ Dor frequency range ($0-8$ \cd\ ) due to the absence of any higher-amplitude frequencies, which would be
indicative of p-mode
pulsation. One-day aliases and harmonics were identified and removed.

\subsection{Spectroscopic Frequency Analysis}\label{spe_freq_gam}

\begin{figure}
\centering
   \includegraphics[trim=0cm 2cm 0cm 0cm,width=0.49\textwidth]{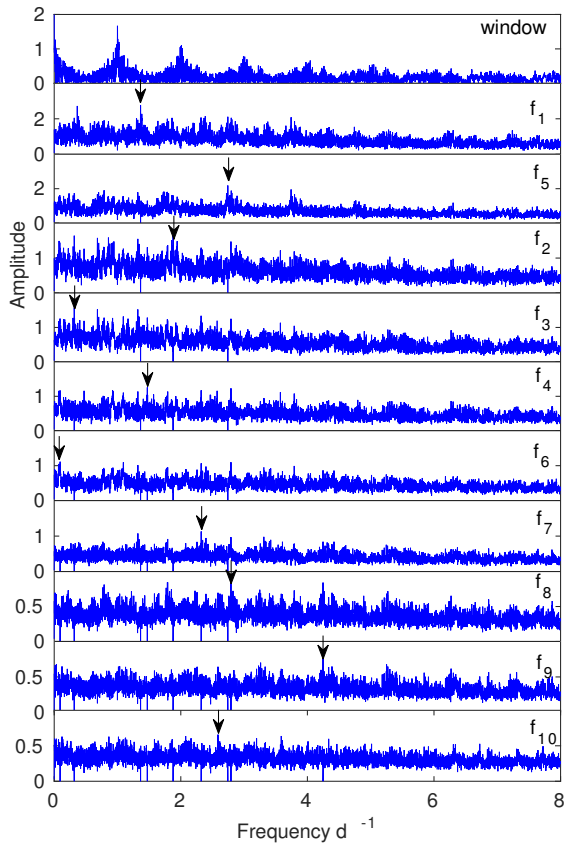}\caption[Fourier spectra showing the \textsc{pbp} identified
frequencies.]{Fourier spectra showing the ten identified frequencies
using
the \textsc{pbp} technique. The top panel shows the spectral window which is representative of the data sampling.}\label{gam_pbpfou}
\end{figure}

We identified ten candidate frequency peaks in the \textsc{pbp} Fourier spectrum. The spectrum is plotted in Figure~\ref{gam_pbpfou} and Table
\ref{gam_freqpbp} details the frequencies found. One-sigma uncertainties are calculated as in \citet{1999DSSN...13...28M}. This provides an underestimate of
the true uncertainties, but can be used as a indicator of the theoretical limitations of the data.
 
\begin{table*}\caption[Frequencies found in the analysis of $\gamma$ Doradus.]{Frequencies found in the analysis of $\gamma$ Doradus. The spectroscopic frequency $f_3$ is identified as a one-day alias of
the $1.32$~\cd\ frequency. The \textsc{pbp} amplitude is measured in normalised intensity units. Sig is the dimensionless spectral significance as described in the text.}\label{gam_freqpbp}

\begin{center}
\begin{tabular}{ccccccccc}
\\
\toprule
&\multicolumn{2}{c}{Spect.}&\multicolumn{3}{c}{Phot.}&\multicolumn{3}{c}{Previous work}\\
\textsc{id} & Freq (\cd) & \textsc{pbp} Amp & Freq (\cd) &Sig& V Amp& Phot.\footnotemark[1](\cd) & Spec.\footnotemark[2](\cd) & Phot.\footnotemark[3](\cd)\\
\midrule
$	f_{1}	$	&	1.3641(2)	&	0.41	&1.36353(2)	&	99	&	1.00	& 1.36354  & 1.365  & 1.36351 \\
$	f_{2}	$	&	1.8783(3)	&	0.25	&1.87828(5)	&	15	&	0.27	&  &  & 1.87  \\
$	f_{3}	$	&	0.3167(2)	&	0.35	& 1.32094(2)	&	109	&	0.96& 1.32098  &  & 1.32093\\
$	f_{4}	$	&	1.4712(3)	&	0.22	& 1.47141(3)	&	46	&	0.51& 1.475  &  & 1.471 \\
$	f_{5}	$	&	2.7428(2)	&	0.33	&\\
$	f_{6}	$	&	0.0934(2)	&	0.28	& 0.91226(5)	&	18	&	0.25\\
$	f_{7}	$	&	2.3222(3)	&	0.22	& &&&\\
$	f_{8}	$	&	2.7924(4)	&	0.16	& &&&\\
$	f_{9}	$	&	4.2471(5)	&	0.15	&&&& \\
$	f_{10}	$	&	2.5876(5)	&	0.13	& 2.56369(5)	&	15	&	0.18&&\\

\bottomrule
\end{tabular}
\\
\footnotemark[1]\footnotesize{
\citet{1994MNRAS.270..905B}}, \footnotemark[2]\footnotesize{\citet{1996MNRAS.281.1315B}},
\footnotemark[3]\footnotesize{\citet{2008AandA...492..167T}}
\end{center}
\end{table*}

We removed alias frequencies from the identifications. This includes $f_7$ which was a $2$\cd\ alias of $f_3$. It is noted that $f_{5}$ is identified as a combination frequency in
\citet{2008AandA...492..167T}, equivalent to the sum of their $f_2 = 1.36351$\cd\ and $f_5 = 1.39$\cd. There was no evidence for a frequency
similar to their $f_5$ in the present data, yet the $f_{5}$ frequency is very close to double the high amplitude frequency $f_{1}$. This means it is
probably a harmonic frequency.

\begin{figure}
\centering
   \includegraphics[width=0.5\textwidth]{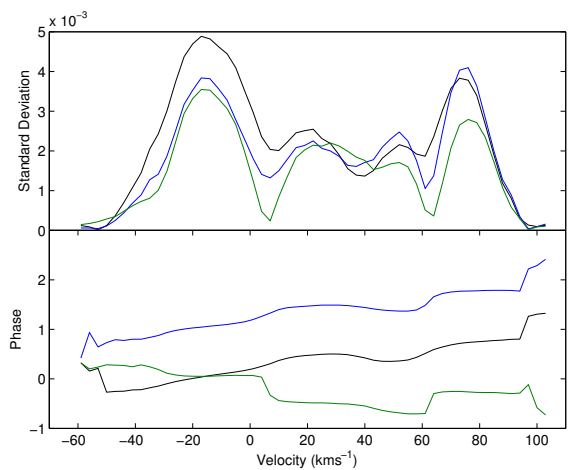}\caption[Standard deviation and phase profiles of
variations of $f_{3}$.]{Comparison of the standard deviation and phase profiles of
$f_{3}=0.3167$ ~\cd\ (black), $f_{3} + 1 $~\cd\ (blue) and $1.32098$~\cd\ (green).}\label{10pbpf4}
\end{figure}

We examined the amplitude and phase diagrams of the \textsc{pbp} identified frequencies in \textsc{famias} to assist in the detection of further alias
frequencies. The frequency $f_{3}$ is close to a one day alias of previously detected photometric frequency $1.32098$~\cd. The standard
deviation profile of $1.32098$~\cd\ is shown with that of the identified $f_{3}$ in Figure~\ref{10pbpf4} and shows a much more symmetric and smooth
variation, as well as much more distinct phase changes. We thus adopted the value of $1.32098$~\cd\ for $f_{3}$ in further analysis.

We noted the two primary frequencies, $f_{1}$ and $f_{3}$ were similar and thus we further examined them to identify any co-dependence. To determine if $f_{3}$ could arise from
the identification of $f_{1}$ in this particular dataset a synthetic dataset was created. The synthetic data were monoperiodic and sampled identically to the observations. As only one frequency was insterted, the Fourier spectrum of these synthetic data shows the frequency convolved with the spectral window of the observations. The removal
of the identified frequency did not show any residual power in this region. The frequency $f_{3}$ is thus unlikely to be an artefact frequency caused by $f_{1}$.

\subsection{Photometric Frequency Analysis}\label{10phofreq}

We analysed a total of $1329$ $V$-band observations from \citet{1992Obs...112...53C}, \citet{1994MNRAS.270..905B}, \citet{1994MNRAS.267..103B} and
\citet{1996MNRAS.281.1315B}. These were used to identify frequencies using a very long timespan of $4785$ days, more than $13$ years. A
observation log of all the
photometry of $\gamma$ Doradus is shown in Table~\ref{2allphot}. Although many of the
datasets collected were multi-colour photometry this section restricts analysis to that of the $V$-filter as this filter
provides the largest number of observations. 

We analysed the $V$-magnitudes of the photometry in SigSpec for frequencies and the results are shown in Table~\ref{gam_freqpbp}.  We discarded alias frequencies. The spectral significance of each frequency is shown in column five of Table~\ref{gam_freqpbp}. Spectral significance is a weighting of the detected frequencies and includes the analysis of the false-alarm probability to remove frequency peaks caused by irregular
data sampling or noise in the data \citep{2007AandA...467.1353R}. We calculated uncertainties in the frequencies as in \citet{2008AandA...481..571K}.
\subsection{Frequency Results}\label{gam_fres}

We extracted the frequencies that occurred in photometry and spectroscopy, giving ten candidate pulsation frequencies. We examined each of these frequencies to combine the results. Frequencies $f_{5}$ and $f_{7}$ have already been discussed as being harmonic and alias frequenquencies, respectively. 

The frequency $f_{6}$ appears in several techniques as either $0.91$~\cd\ or the $1-f$ alias, $0.09$~\cd. The shape of the standard deviation and phase appears reasonable for a pulsation but has
asymmetric amplitudes in the standard deviation profile. Regardless of the shape, a frequency lower than about
$0.3$~\cd\ is not expected to be found in a $\gamma$ Dor star unless a mode is retrograde, so we also examined the one-day alias frequencies $1.0934$~\cd and $0.9123$~\cd.
Both the frequencies had reasonable standard deviation profiles so this frequency remained a candidate for further analysis.

Frequencies $f_8$, $f_9$ and $f_{10}$ have low amplitudes, unsymmetric standard deviation profiles and likely arise from combinations of lower frequencies (e.g. $f_8=f_3+f_4$) so were not considered further.

To prepare the frequencies for mode-identification, we calculated a least squares fit of all the frequencies together. This process dramatically improved the symmetry of the standard
deviation profiles of the frequencies $f_1$ to $f_4$. The last frequency, $f_6$ was
however severely distorted for both the $0.9123$~\cd\ and the $1.0934$~\cd\ frequencies. This led us to the conclusion that the frequency, although having a reasonable
standard deviation profile on its own, must be closely related on
one or more of $f_1$ to $f_4$ and thus not independent. We therefore did not consider it further in the mode identification and discounted it as a pulsation
frequency. 

We checked for any characteristic frequency or period spacings between the frequencies $f_1$ to $f_4$. No common spacings could be attributed to the
physical properties of the star.

We performed mode identification of the first four frequencies using the spectroscopic identifications for $f_1$, $f_2$ and $f_4$ and the photometric identification
of the value for $f_3$ due to the improved standard deviation profile.

\section{Mode Identification}\label{gam_modeid}

The spectroscopic mode identification showed multiple well-fitting modes. This led us to use the photometric data to constrain $l$ and thus differentiate the best modes in
spectroscopy. The photometric mode identification is discussed first in Section \ref{gam_modeid_ph} and then informs the spectroscopic analysis in Section
\ref{gam_modeid_sp}.

\subsection{Photometric Mode Identification}\label{gam_modeid_ph}

The largest multi-colour photometry set of the full photometric dataset was taken in the Str\"omgren filters in 1993 and 1994. The two observing runs yielded a total of $428$ observations in the four filters. 

We used \textsc{famias} to identify or constrain the modes of the frequencies $f_{1}$ to $f_{4}$. We chose stellar parameters of
$\Teff = 7100
\pm
150$ K
and $\logg = 4.0 \pm 0.2$ as being representative of most of the values found in the literature (see Table~\ref{10param}). The mass was
modelled using $1.8$~\msun\ using the non-adiabatic Warsaw-New Jersey/Dziembowski code\footnote{Grids of atmospheric parameters have been computed by Leszek
Kowalczuk and Jagoda Daszynska-Daszkiewicz
(\url{http://helas.astro.uni.wroc.pl/deliverables.php}) using Kurucz and NEMO atmospheres.
Pulsational grid for main-sequence stars with 1.8 to 12 M(sun) computed by Jagoda Daszynska-Daszkiewicz, Alosha Pamyatnykh, and Tomasz Zdravkov
using the non-adiabatic Warsaw-New Jersey/Dziembowski code.
\url{http://helas.astro.uni.wroc.pl/deliverables.php?active=opalmodel&lang=en}}. The metallicity was modelled as solar and the microturbulance calculated as
$2$~\kms. Both the
amplitude ratios and the phase differences were tested for $l = 1$ to $l = 5$ for all the four frequencies. The results are summarised in Table
\ref{gam_phot_mode}.

\begin{figure}
\centering
   \includegraphics[width=0.5\textwidth]{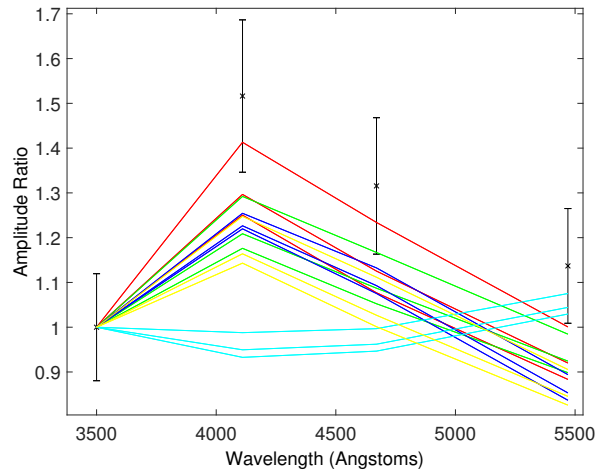}\caption[Amplitude ratios for the mode identification for $f_1$]{Amplitude
ratios for the mode identification of $f_1$. The colours show the models of different degree, specifically $l~=~1$ (red), $l~=~2$ (green), $l~=~3$
(blue), $l~=~4$ (cyan), $l~=~5$ (yellow). Although a unique solution is not obtained, the results favour the $l~=~1$ mode.}\label{gam_amrat}
\end{figure}

The closest fits for the amplitude ratios of $f_{1}$ to $f_{4}$ were to $l~=~1$. An example of the amplitude ratio fit to $f_1$ is given in Figure
\ref{gam_amrat}. The phase differences place almost no restrictions on the modes. These results significantly constrain the spectroscopic mode
identification in the following section.

\begin{table}\caption[Mode degrees, $l$, found in the photometric
filters.]{Mode degrees, $l$, that fit the amplitude ratios and phase differences of the Str\"omgren photometric
filters.}\label{gam_phot_mode}
\begin{center}
\begin{tabular}{lllll}
\toprule
Filter & $f_1$ & $f_2$ & $f_3$ & $f_4$\\
\midrule
\multicolumn{5}{c}{Amplitude Ratio}\\
$A(v)/A(u)$ &  1 &  all & - &-\\
$A(b)/A(u)$ &  1 &  all & - &-\\
$A(y)/A(u) $ & 4  &  all &  4 &4\\
\\
\multicolumn{5}{c}{Phase Difference}\\
$P(v - u)$ & all & all & all &1,3,4,5\\
$P(b - u)$ & all & all & all &all\\
$P(y - u)$ & all & all & all &all\\
\bottomrule
\end{tabular}
\end{center}
\end{table}
  
\subsection{Spectroscopic Mode Identification}\label{gam_modeid_sp}

We modelled the frequencies found in \textsc{famias} to determine the best fitting mode. It was noted in the analysis that
the standard deviation profiles were shifted such that the centre of the
symmetric deviations was translated from the centre of the line profile. The asymmetry
of the pulsational distortion of the line profile results in poorer fits for the modes but is not expected to affect the overall identification. This restricts the identification to only using the Amplitude and Phase
(\textsc{ap}) fits and
not considering the Zero-point, Amplitude and Phase (\textsc{zap}) fit in \textsc{famias}. This is usual practice for the fitting of g-modes, as the zero-point
profile fit
can statistically dominate over the standard deviation and phase fits and is generally poorer at distinguishing modes.

\subsubsection{Line Profile Parameters}

\begin{table}\caption[Line-profile parameters found in the zero-point fit.]{Line-profile parameters found in the zero-point fit. The minimum $\chi^2$ value was
$489$. Non-physical values of radius and mass are discussed in the text.}\label{zpparamgam}
\begin{center}
\begin{tabular}{ccccccc}
\toprule
$\chi^2$ & Radius & Mass & $i$ & \vsini&Equiv.& avg vel. \\
 & (\rsun) & (\msun) & ($\degree$) & (\kms) & Width.& (\kms)\\
\midrule
489 & 7.51 & 2.37 & 9.02 & 56.6 ($\pm$ 0.5) & 8.94 & 22.9\\
\bottomrule
\end{tabular}
\end{center}
\end{table}

We fitted the zero-point profile to measure \vsini, equivalent width and velocity offset. The parameters mass, radius and
inclination were also permitted to vary. The values for mass and radius are used to determine the
horizontal-to-vertical amplitude ratio, $k$, and the values do not otherwise affect the mode-identification. Unrealistic values for mass and radius can arise when fitting a g-mode, which has mostly horizontal
motion, with models designed for p-modes, which have mostly vertical motion. The geometric shape of the pulsation $(l,m)$ is independent of amplitude in the fits. 

The best fit model had a $\chi^2$ of $489$. It is expected that zero-point fits will have high $\chi^2$ values as they have the lowest measurement
uncertainties, thus requiring more precise $\chi^2$ fits for a lower $\chi^2$ value. These uncertainties do not reflect all the natural uncertainties in the
acquisition and reduction processes but are suitable to be used to judge
the best fit profile. The model found with the best-fit
values for the parameters is detailed in Table~\ref{zpparamgam}. 

From the zero-point fit we derived a \vsini\ of $56.6~\pm~0.5$~\kms\ . The
uncertainties are from a $95\%$ confidence limit calculated from the critical values of
the $\chi^2$ distribution added to the minimum $\chi^2$ value. 

\begin{figure}
\centering
\subfigure[$f_{1}=1.3641$~\cd\ with a $(1,1)$ mode ($\chi^{2}=10.6$).]{  
\includegraphics[width=0.4\textwidth, trim=0.5cm 0.3cm 1cm 0.6cm, clip=true]{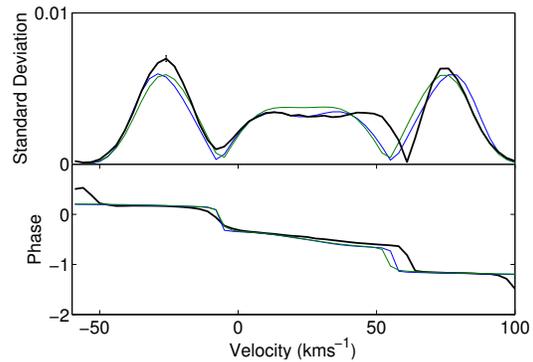}\label{10mode2f1}
}
\subfigure[$f_{2}=1.8783$~\cd\ with a $(1,1)$ mode ($\chi^{2}=3.88$).]{  
\includegraphics[width=0.4\textwidth, trim=1cm 0.3cm 1cm 0.6cm, clip=true]{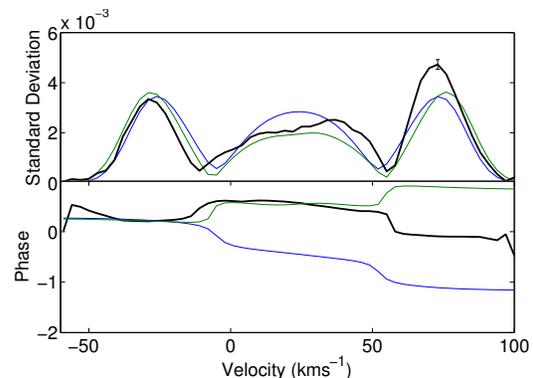}\label{10mode2f2}
}
\subfigure[$f_{3}=1.3209$~~\cd\ with a $(1,1)$ mode ($\chi^{2}=11.8$).]{  
\includegraphics[width=0.4\textwidth, trim=1cm 0.3cm 1cm 0.6cm, clip=true]{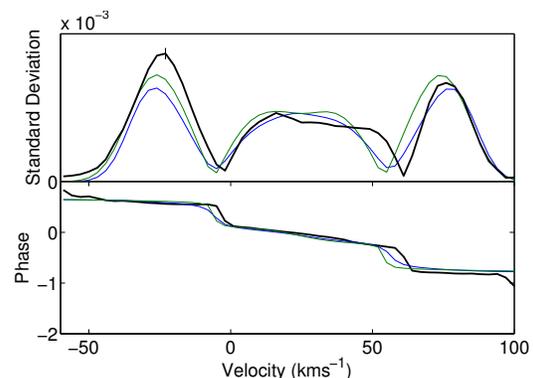}\label{10mode2f4}
}
\subfigure[$f_{4}=1.4742$~\cd\ with a $(2,0)$ mode (blue, $\chi^{2}=2.6$) and a $(2,-2)$ mode (green, $\chi^{2}=6.4$).]{  
\includegraphics[width=0.4\textwidth, trim=1cm 0.3cm 1cm 0.6cm, clip=true]{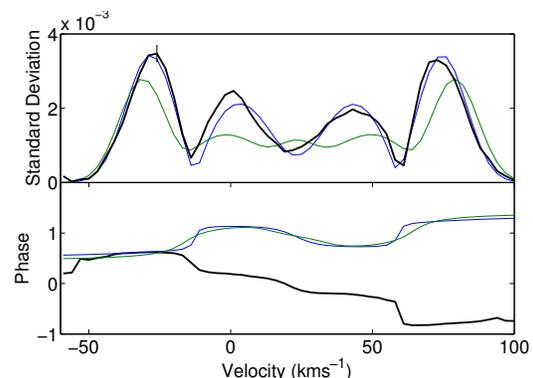}\label{10mode2f3}
}

\caption[Mode identification of the four identified frequencies.]{Mode identification of the four identified frequencies. The standard deviation and phase profiles (black) with the individual fits (blue) and combined
fit (green) are shown for the four identified frequencies. The combined fit has a $\chi^2$ of $14.2$. An indication of the uncertainty is shown by
the error bar.}\label{10modeidfig}
\end{figure}

\subsubsection{Individual Mode Identification}
\begin{table*}
\caption[Best fit parameters for a multi-mode parameter search.]{Best fit parameters for a multi-mode parameter search using $f_1$ to $f_3$ and allowing the mode}\label{10fcombid1}
of $f_4$ to vary. Velocity amplitudes (Vel.) are in units of \kms. Models with $\chi^2$
less than $25$ are shown, with other mode fits having much higher $\chi^2$ ($>50$).

\begin{center}
\begin{tabular}{cccccc|cc|cc|cc|cccc}
\toprule
 & & & & & &\multicolumn{2}{c}{$f_1$}& \multicolumn{2}{c}{$f_2$}&\multicolumn{2}{c}{$f_3$}&\multicolumn{4}{c}{$f_4$}\\
$\chi^2$	&	$R$	&	$M$	&	$i$	&	\vsini	&	avg. Vel.  &	Vel.	&
Ph. &	Vel.	&	Ph.	&	Vel.	&	Ph.		&	Vel.
&	Ph.	&	$l_{f4}$	&	$m_{f4}$\\
	&	(\rsun)	&	(\msun)	&	($\degree$)	&	(\kms)	&	(\kms)	&		&
 &		&		&	&		&		&		&	
&		\\
\midrule
14.2	&	3.87	&	5.21	&	31.9	&	58.4	&	24	&	2	&	0.76	&	2.7	&	0.8	&	3.86	&	0.68	&1.3	&	0.19	&	2	&	-2	\\
																														
14.8	&	4.31	&	7.23	&	30	&	59.9	&	24	&	2	&	0.76	&	2.7	&	0.8	&	1.03	&	0.17	&1.3	&	0.18	&	1	&	1	\\
																														
18.6	&	3.87	&	5.21	&	30	&	59.6	&	24.7	&	2	&	0.75	&	2.7	&	0.8	&	0.01	&	0.2	&1.3	&	0.17	&	2	&	2	\\
																														
20.3	&	4.39	&	7.91	&	30	&	59.6	&	24	&	2	&	0.76	&	2.7	&	0.8	&	0.8	&	0.68	&1.3	&	0.19	&	3	&	-2	\\
																														
21.5	&	4.17	&	6.48	&	30	&	59.4	&	24.7	&	2	&	0.75	&	2.7	&	0.8	&	0.4	&	0.66	&1.3	&	0.18	&	2	&	-1	\\
																														
21.5	&	4.24	&	6.86	&	30	&	59.9	&	24	&	2	&	0.76	&	2.7	&	0.8	&	1.35	&	0.62	&1.3	&	0.18	&	0	&	0	\\
																														
22.1	&	4.09	&	6.19	&	30	&	59.2	&	24.7	&	2	&	0.75	&	2.7	&	0.8	&	0.25	&	0.91	&1.3	&	0.18	&	3	&	-3	\\
																														
22.3	&	4.39	&	7.68	&	30	&	59.6	&	24	&	2	&	0.76	&	2.7	&	0.8	&	0.48	&	0.51	&1.3	&	0.18	&	1	&	-1	\\

\bottomrule
\end{tabular}
\end{center}
\end{table*}
We performed mode identification on each of the frequencies $f_1$ to $f_4$ individually. This is done to constrain the parameter space for the simultaneous
mode fit. We found a $(1,1)$ mode best fit frequencies $f_1$, $f_2$ and $f_3$, but $f_4$ was best fit using a $(1,1)$,$(2,0)$ or $(2,-2)$ mode (Figure \ref{10modeidfig}). To further restrict the parameter space to physical regions, we investigated the rotation and pulsation limits of the models.

We trialled the frequencies identified for
typical $\gamma$ Dor parameters, M$~=~1.6$~\msun\ and R$~=~1.5$~\rsun, which showed that all the modelled combinations were physically possible except for
$(1,1)$ and $(2,2)$ modes identified for $f_{3}$. For the $(1,1)$ mode, the minimum possible inclination of the star which produces a real frequency is
$40\degree$. The $(2,2)$ mode, on the other hand, is not physically possible for any value of $i$ and is excluded from future consideration for this frequency.

\textsc{famias} has a general operating limit of $\frac{f_{rot}}{f_{pul}}~<~0.5$. Fourteen models had individual fits that were higher than this limit.
However, none of them were excluded on this principle alone. Most are only slightly greater than $0.5$ and low values of $m$
are less affected by the rotational limit. Lower inclinations increase the $\frac{f_{rot}}{f_{pul}}$ ratio. For most of the above modes, restricting the inclination removes the risk of producing non-physical models.

We used typical solutions for \vsini\ and inclination from the individual mode fits, and mass and radius as described above, to determine an estimate of the equatorial
rotational velocity, rotation frequency and critical limits of \vsini\ and inclination when stellar break-up occurs.
A \vsini$~\approx~57$~\kms\ and $i~\approx~60\degree$ corresponds to a star with an equatorial rotation velocity of around $66$~\kms, and rotational
frequency of approximately
$0.9$~\cd. To exceed break-up velocity the critical \vsini\ is $390$~\kms\ and critical minimum inclination of $7\degree$. Models with inclination of less than $7\degree$ were thus rejected. The star is not otherwise close to rotating
near the break-up velocity. A rotational frequency near $1$~\cd\ is expected for $\gamma$ Doradus given the \vsini. These values range from around $1.5$~\cd\ to $0.75$~\cd for inclinations in the range of $30\degree-90\degree$. 

\subsubsection{Combined Mode Identification}
\begin{figure*}
\centering
\subfigure{  
\includegraphics[width=0.43\textwidth]{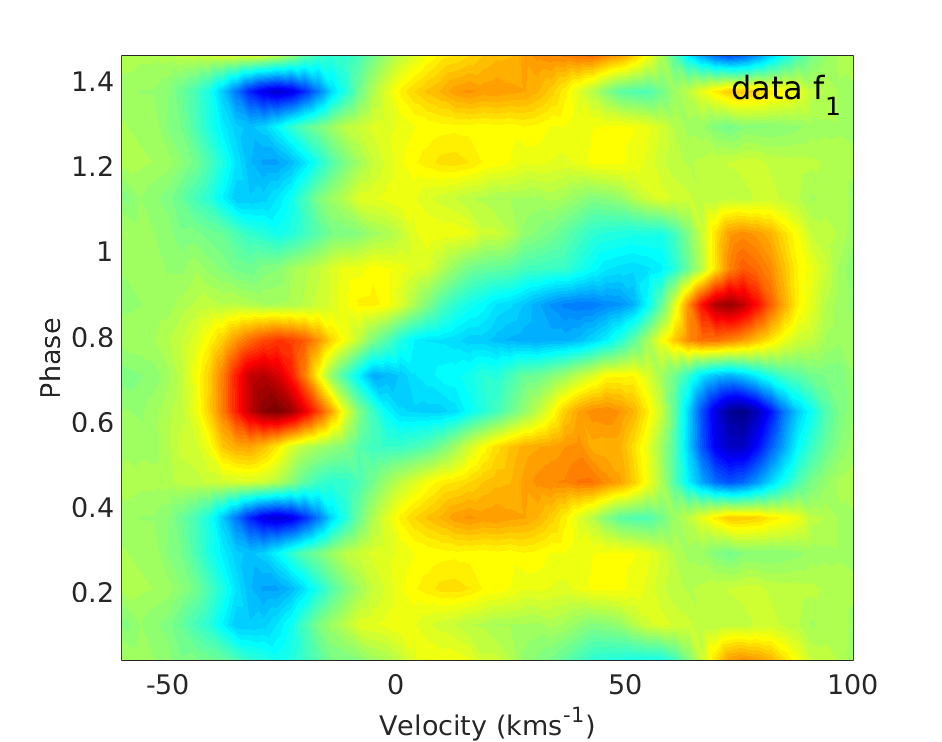}\label{10phdat1}
}
\subfigure{  
\includegraphics[width=0.43\textwidth]{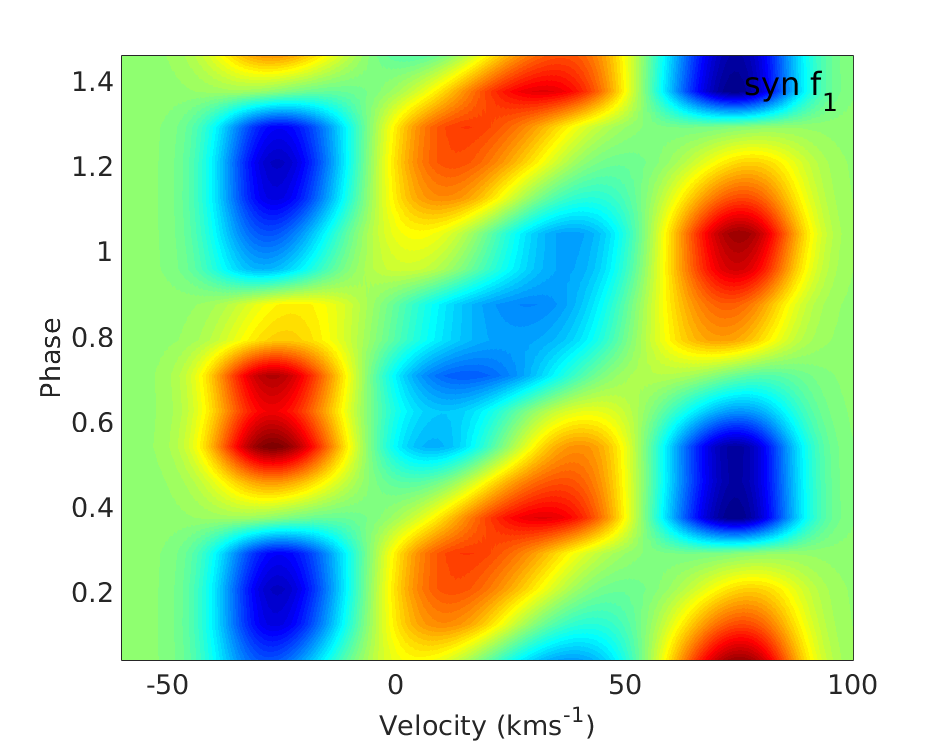}\label{10syndat1}
}
\subfigure{  
\includegraphics[width=0.43\textwidth]{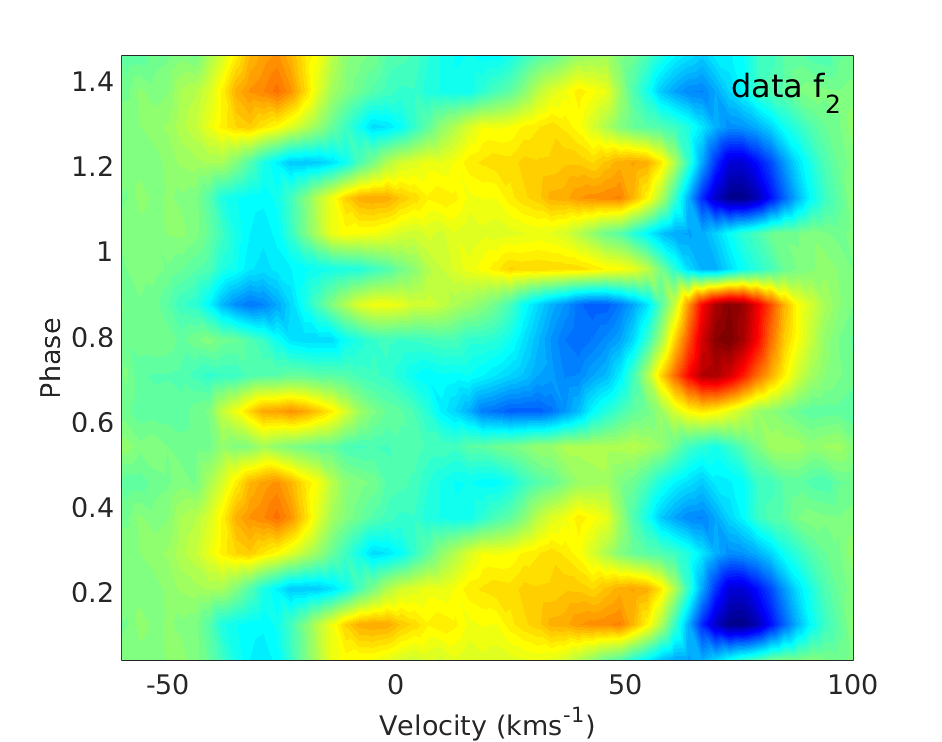}\label{10phdat2}
}
\subfigure{  
\includegraphics[width=0.43\textwidth]{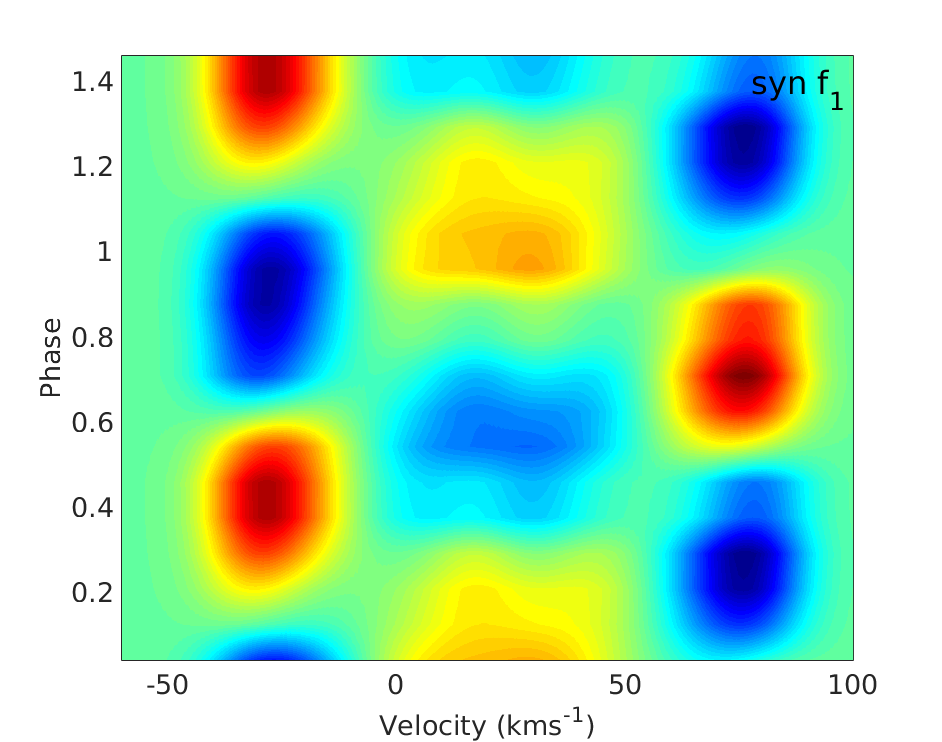}\label{10syndat2}
}
\subfigure{  
\includegraphics[width=0.43\textwidth]{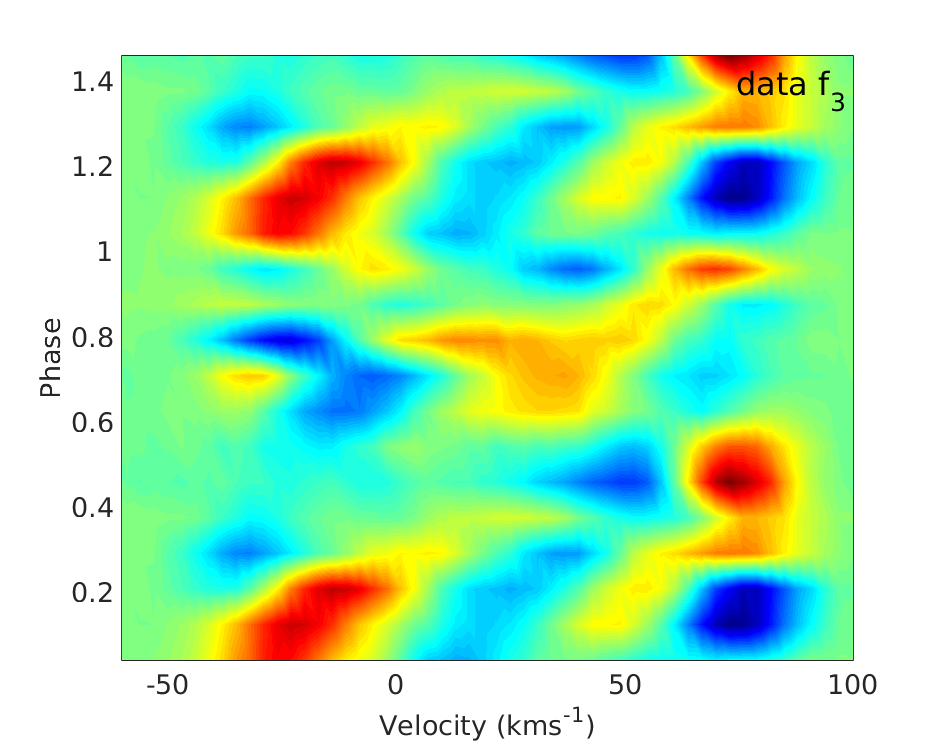}\label{10phdat4}
}
\subfigure{  
\includegraphics[width=0.43\textwidth]{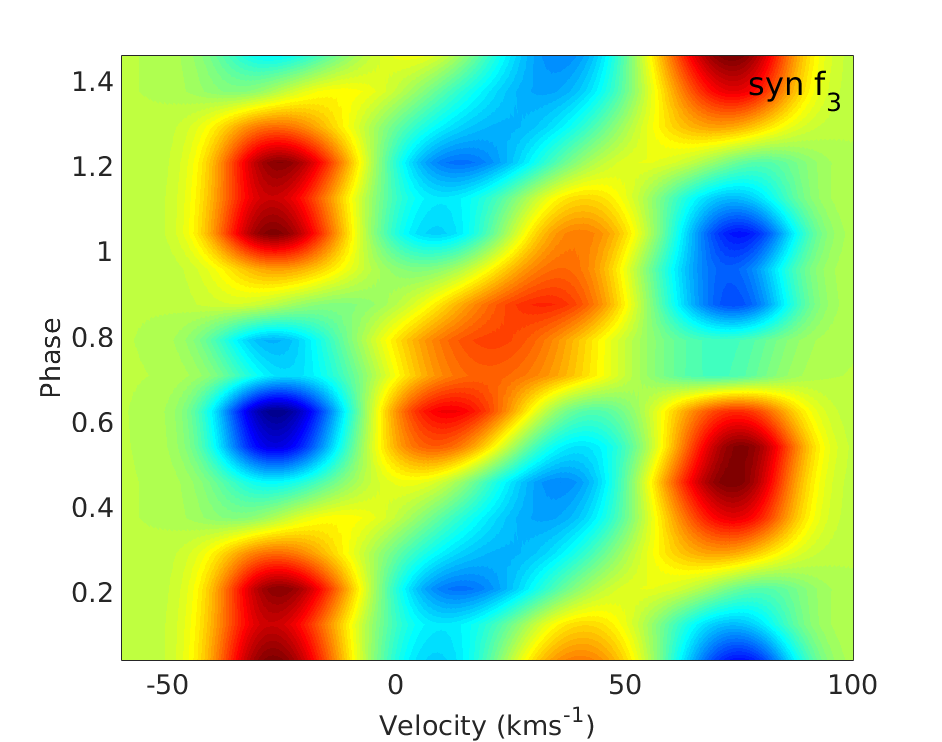}\label{10syndat4}
}
\caption[Residual line-profile variations of the spectra and models for $f_1$, $f_2$ and $f_3$.]{Phased residual line profile variations of the observed spectra
(left column) and the synthetic models using the combined fit (right column) for
frequencies $f_1$, $f_2$ and $f_3$, all fitted with $(1,1)$ modes.}\label{phassyn}
\end{figure*}

\begin{figure}
\centering
\subfigure{  
\includegraphics[width=0.43\textwidth]{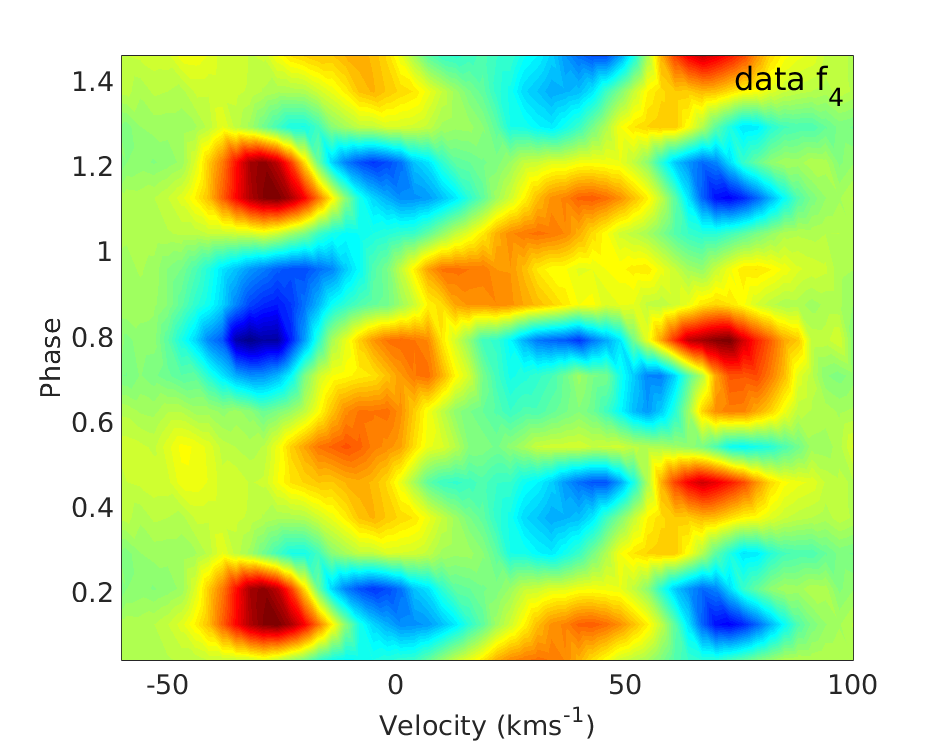}\label{10phdat1b}
}
\subfigure{  
\includegraphics[width=0.43\textwidth]{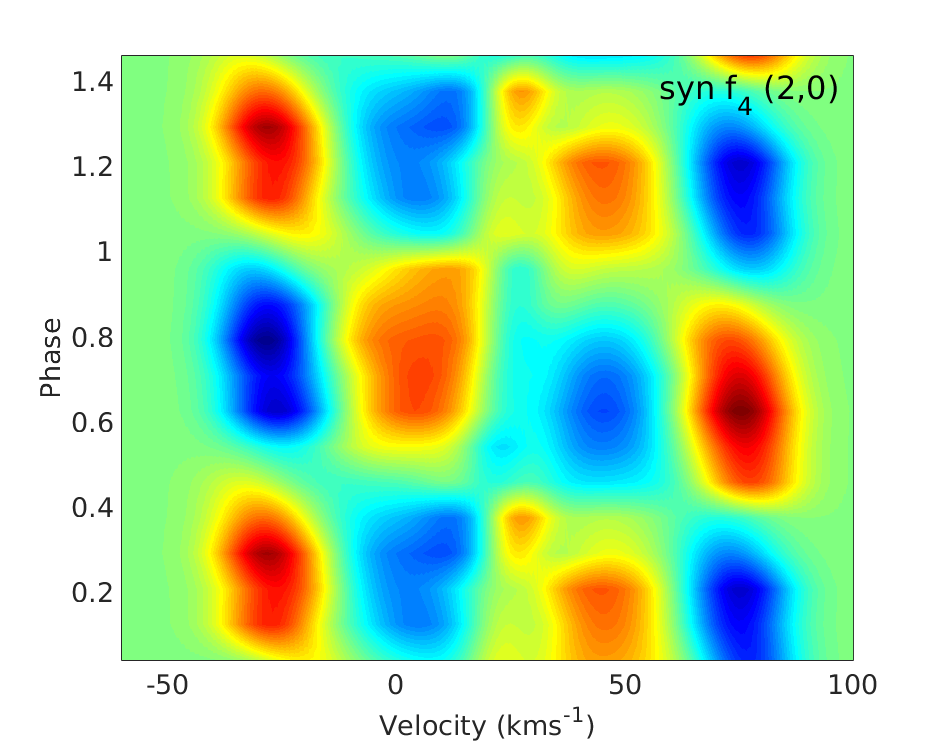}\label{10syndat1b}
}
\subfigure{  
\includegraphics[width=0.43\textwidth]{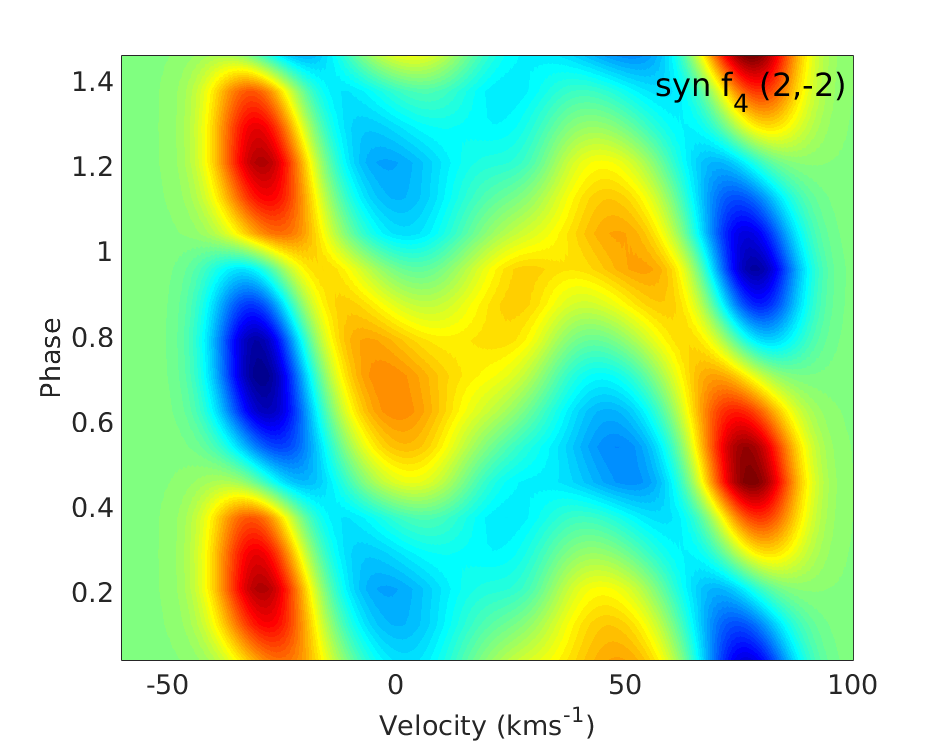}\label{10phdat2b}
}
\caption[Residual line-profile variations of the spectra and models for $f_4$.]{Phased residual line profile variations of the observed spectra
for $f_4$ and two models. The first model is the individual
frequency fit with mode $(l,m)~=~(2,0)$ and the second model is the combined mode fit with a $(2,-2)$ mode.}\label{f3smells}
\end{figure}

The photometric mode identification suggests a best fit $l$ value for all the frequencies of $l~=~1$. The identification of the $f_1$, $f_2$ and $f_3$ frequencies can be therefore concluded to be $(1,1)$ modes from the best fits in the individual mode analysis. The identification of $f_4$ is more complex. The mode $(2,2)$ must first be excluded as it has already been deemed non-physical. The shape of the $(1,1)$ mode is also excluded as the best fit model still displays a three-bump structure
that does not fit the four bumps
observed for this frequency. It is not expected that $f_4$ should have a much higher velocity amplitude than the other frequencies, thus the identification
of the above modes suggests that modes $(2,0)$, $(3,0)$ and $(2,-2)$ should be favoured. Since the mass and
radius of the star are allowed to vary non-physically, it is prudent to not discard the modes based on the amplitudes alone. The parameter space
searched for $f_4$ included the higher amplitude frequencies with all models with $l~=~2$ and $l~=~3$ considered so as not to miss any possible identifications.

We constrained the inclination to be a minimum of $30\degree$. Inclinations
lower than this limit produce very high amplitude intrinsic pulsation frequencies, which are not theoretically predicted for self-excited g-modes in these stars. Additionally, the observational
amplitudes of the pulsations and the probability of the existence of an equatorial wave-guide for a rotating, pulsating star make it
unlikely that such amplitudes of pulsation could be observed in spectroscopy, and would be nearly impossible to detect in photometry. 

We undertook a combined mode-identification search to fit the stellar parameters and to find the best mode fit for $f_4$ that was consistent with the other
frequencies. The best fit models are shown in Table~\ref{10fcombid1} and
plotted in Figure~\ref{10modeidfig}. Not all the fits in the table should be regarded as accurate fits to the
star. The radius and mass parameters have been allowed to vary considerably beyond the physical limits for a $\gamma$ Dor star in order to produce visible
amplitudes of pulsation. Even with these large values, the velocity amplitudes for most of the fits are much higher than expected.

From these models, and the fits, it is clear that the three frequencies, $f_1$, $f_2$ and $f_3$, are well modelled by
$(1,1)$ modes. The other
frequency, $f_4$, does not appear to be well modelled by the lowest $\chi^2$ fit. To see if the fits matched well over the entire phase of each pulsation frequency, we computed synthetic profiles of the frequencies using the
modes and parameters determined in the best fit in Table~\ref{10fcombid1}. These are compared with the phased line profiles observed in Figure
\ref{phassyn} for $f_1$, $f_2$ and $f_3$ and Figure~\ref{f3smells} for $f_4$. The $(1,1)$ mode frequencies in Figure~\ref{phassyn} were all well fitted in phase
space and this supports their identification. The two proposed fits for $f_4$ in Figure~\ref{f3smells} are less clear, but it can be seen that the $(2,0)$ mode
better describes the variation over the phase of the frequency, even though it is inconsistent with the best fits for the combined mode identification.

We calculated the co-rotating frequencies of the identifications $f_1$ to $f_4$ to identify if the $l~=~2$ identification for $f_4$ was consistent with the
expectation that higher order modes have higher frequencies \citep{2005AandA...435..927D}. The co-rotating frequencies were calculated as in \citet{1992ApJ...394..670D}, $f_{\mathrm{corot}}=f_{\mathrm{obs}}-m(1-C)f_{\mathrm{rot}}$, using the parameters of the best fit in Table~\ref{10fcombid1} and $C=\frac{1}{l(l+1)}$.

 We found the frequencies $f_1$, $f_2$ and $f_3$ to have co-rotating frequencies ranging from $1.0388$~\cd\ to $1.5962$~\cd. The
co-rotating frequency for $f_4$ is $1.4712$~\cd\ and $2.4115$~\cd\ for the $(2,0)$ and $(2,-2)$ mode respectively. The $(2,-2)$ mode frequency is higher
than the $l~=~1$ identifications so this supports the identification of $f_4$ as a $(2,-2)$ mode, but as these co-rotating frequencies can only be estimated this does not rule out the $(2,0)$ mode. 

Due to the poorness of the fit obtained for $f_4$, we also performed the multi-mode fit on just the well-identified frequencies
$f_1$, $f_2$ and $f_3$. This produced best fit parameters nearly identical to the combined mode identification with a $\chi^2$ of $14.0$. The difficulties in
fitting the four modes simultaneously questions the validity of the $f_4$ frequency, but there is no other evidence for excluding it as a valid frequency
arising from a $(2,0)$ mode. Parameters usually reported from these fits include the \vsini\ and the inclination. Fitting the zero-point profile above gave a credible
\vsini\ and this was not seen to vary significantly in the individual or combined mode fits. Although the best fit inclination was found to be at $31\degree$, this would need to be confirmed using a model that accounts for all the observed frequencies and modes. Because of the difficulties with the identification
of the modes in this star this would likely have to be a model that accounts for 
the rotation and horizontally-dominated motion of a $\gamma$ Dor pulsation.

\section{Discussion}\label{disc4}

The implications of fitting a non-physical mass and radius in the model are concerning. The mass and radius, along with the other parameters, are all allowed to vary in the individual
and combined
mode fits. The high values for the velocity amplitudes are an indication of a non-physical fit for a $\gamma$ Dor star. The amplitude of pulsation alone does not have
a significant effect on the shape of the line profile. In fact, changing the mass and radius has a much more significant effect on
the observed amplitude than the intrinsic velocity amplitude, since this affects the value of $k$, the horizontal-to-vertical amplitude ratio.

The frequencies found in the analysis of $\gamma$ Doradus are almost identical to those found in previous spectroscopic and photometric studies. No new frequencies were found but
this work confirms the proposal of the frequency at $1.878~$\cd\ by \citet{2008AandA...492..167T}. The frequencies $f_1$ and $f_3$ are shown to be
stable over twenty years since their first identification by \citet{1992Obs...112...53C}. This long-term stability of frequencies is also seen in the
$\gamma$ Dor star HD\ 108100, \citep{1997AandA...324..566B}. 

$\gamma$ Doradus shows an excellent agreement between the frequencies found in photometry and spectroscopy. This has not always been the case for previously studied
$\gamma$ Dor
stars (e.g. \citeauthor{2008AandA...489.1213U}, \citeyear{2008AandA...489.1213U}, \citeauthor{2015MNRAS.447.2970B}, \citeyear{2015MNRAS.447.2970B}) although other stars show similar agreement (e.g. \citeauthor{2012arXiv1209.6081B}, \citeyear{2012arXiv1209.6081B}, \citeauthor{2014AandA...568A.106S}, \citeyear{2014AandA...568A.106S}). It is not known why such differences should occur, but this work continues with the complementary data from large photometric surveys and targeted high-resolution spectroscopic campaigns. The power of combining the techniques lies in the precise identification of frequencies from photometry with the two-dimensional information provided from the line profiles. 

An investigation into the aliases and combination frequencies led us to remove of the frequency at $2.74$~\cd\ as a harmonic of $f_1$.
\citet{2008AandA...492..167T} also removed this frequency, identifying it as a combination of $1.36351$~\cd\ and their proposed $1.39$~\cd. We found no
evidence for a further frequency near $1.39$~\cd\ in this analysis. The relationship between the two close frequencies $f_1$ and $f_4$ was
examined but no evidence for a link between them was discovered. 

\citet{1996MNRAS.281.1315B} found modes of $(l,m)~=(1,1)$, $(3,3)$ and $(1,1)$ for the frequencies $f_1$, $f_3$ and $f_4$ respectively, using spectroscopic
line profile variations with further assumptions. Additionally, \citet{Dupret2005} found $f_{1}$ and $f_{3}$ to be both well-fitted by $l=1$ modes. These results confirm the
identification of $(l,m)~=(1,1)$ for $f_{1}$, $f_2$  and $f_{3}$. They also give independent confirmation of the robustness of the g-modes detected in
\textsc{famias}, even when the strong horizontal nature of a g-mode pulsation forces the use of non-physical parameters to obtain the fit. The mode
identification for $f_4$ showed a best fit with a $(2,0)$ mode but was not consistent with the fits of the other frequencies. Despite this result, the frequency
is not discarded as a pulsation frequency. The co-rotating frequencies of the modes found support the identification of $f_4$ as a higher degree mode than the
other frequencies. It is hoped that, with more physically descriptive models of $\gamma$ Dor stars, this inconsistency will be
resolved.

We determined stellar parameters of \vsini\ and inclination in this analysis. The measurement of the zero-point line profile gave a value of $56.6~\pm~0.5$~\kms\
based on a $95\%$ confidence interval. Inclination was best fit to be $31\degree$ for a multi-modal fit of the four frequencies. However, this parameter is
often not well constrained by mode identification. This compares to the values of \citet{1996MNRAS.281.1315B} and \citet{2013ApJ...762...52B} who
found inclination of $\approx70\degree$. The
existence of a debris disc around $\gamma$ Doradus \citep{2010ApJ...710L..26K} may help to improve pulsation models by constraining the inclination of the star
(assuming the rotation and pulsation axes align with the disc axis). The absence of evidence of disc signatures in the spectra of $\gamma$ Doradus suggests the
disc is sparse across the stellar surface, or potentially does not cross the line-of-sight to the star. 

The Transiting Exoplanet Survey Satellite (TESS), expected to be launched in March 2018, will perform a two-year high-precision time-resolved photometric survey of nearly the entire sky. $\gamma$ Doradus is proposed\footnote{https://tasoc.dk/info/targetlist.php} to be observed in short-cadence (2 minutes) and is observable for three epochs \citep{tvguide} of around 27 days in length. This ultra-precise dataset will be provide valuable precision photometric data to align with spectroscopic results.
\FloatBarrier
\section{Conclusion}\label{10disc}
The study of $\gamma$ Doradus as the prototype of its class, is important to furthering current understanding of these pulsations. The star's visual brightness
and southerly latitude has made it an ideal target for southern hemisphere observers. The multitude of photometric results also allows confirmation of
photometric and spectroscopic techniques for these stars. Despite the requirement to go beyond physically possible parameter space in modelling the mass and
radius of the star, the mode identification was successful and proves the validity of this technique. It is hoped that using non-physical models will not be
necessary with improvements to theoretical models, which will be able to account for rotational effects and be specifically adapted to identify
g-modes. The star $\gamma$ Doradus will be an excellent candidate for testing such models.

\section{Acknowledgements}
This work was supported by the Marsden Fund administered by the Royal Society of New Zealand.

The authors acknowledge the assistance of staff at the University of
Canterbury's Mt John Observatory.

We appreciate the time allocated at other facilities for multi-site campaigns and the numerous observers who make acquisition of large datasets possible.

This research has made use
of the {\sevensize SIMBAD} astronomical database operated at the CDS in
Strasbourg, France.

Mode identification results obtained with the software package {\sevensize FAMIAS} developed
in the framework of the FP6 European Coordination action {\sevensize HELAS}
(http://www.helas-eu.org/).

The authors wish to thank L. A. Balona for providing data and discussions regarding this paper.

We are thankful to Simon J. Murphy for helpful
comments which improved this manuscript.
\label{lastpage}
\bibliography{references}{}
\bibliographystyle{mn2e}
\end{document}